\documentclass[aps,prb,twocolumn,showpacs,preprintnumbers,amsmath,amssymb]{revtex4-1}
\usepackage{graphicx}
\usepackage{color}
\begin{document}
\setcounter{page}{1}
\title[]{O-vacancy as the origin of negative bias illumination stress instability in amorphous
In-Ga-Zn-O thin film transistors }

\author{Byungki Ryu, Hyeon-Kyun Noh, Eun-Ae Choi, and
K. J. Chang\renewcommand{\thefootnote}{\alph{footnote})}
\footnote{Electronic mail: kchang@kaist.ac.kr}}
\affiliation{Department of Physics, Korea Advanced Institute of Science and
Technology, Daejeon 305-701, Korea}
\date{\today}

\begin{abstract}
We find that O-vacancy ($V_{\rm O}$) acts as a hole trap and plays a role in negative bias
illumination stress instability in amorphous In-Ga-Zn-O thin film transistors.
Photo-excited holes drift toward the channel/dielectric interface due to small
potential barriers and can be captured by $V_{\rm O}$ in the dielectrics.
While some of $V_{\rm O}^{+2}$ defects are very stable at room temperature, their original
deep states are recovered via electron capture upon annealing.
We also find that $V_{\rm O}^{+2}$ can diffuse in amorphous phase, inducing hole
accumulation near the interface under negative gate bias.
\end{abstract}


\maketitle

Transparent amorphous oxide semiconductors such as amorphous indium-gallium-zinc
oxide ($a$-IGZO) have attracted much attention because of their use as an active channel in
transparent thin film transistors (TFTs).\cite{K.Nomura:Nature04}
Oxide TFTs have shown superior performance, exhibiting very high field-effect mobilities,
compared with amorphous Si TFTs.\cite{K.Nomura:Nature04, T.Kamiya:JDT09}
However, the instability of oxide TFTs remains as an obstacle to overcome for practical applications
to electronic devices.

The device stability is sensitive to bias/current stress, temperature, light illumination, and water vapor
exposure.\cite{A.Suresh:APL08, J.-M.Lee:APL09, K.Nomura:APL09, M.E.Lopes:APL09, J.Lee:APL09, ZTO PPC,
Fung:JID, Takechi:JJAP09, K.Lee:APL09, ETRI-J.09}
Under positive gate bias stress, oxide TFTs exhibited positive shifts of the threshold voltage ($V_{\rm th}$),
which were attributed to electron traps in gate dielectrics and/or interfaces.
\cite{A.Suresh:APL08, J.-M.Lee:APL09, K.Nomura:APL09,  M.E.Lopes:APL09, J.Lee:APL09}
Devices usually maintain good stability under negative bias stress.
On the other hand, the instabilities of the TFTs under negative bias illumination stress (NBIS) are significant,
with negative shifts of $V_{\rm th}$.
In amorphous zinc-tin oxide TFTs, a shift of $V_{\rm th}$ of less than 2 V was observed upon
illumination with visible light, with the recovery time of about 20 hours,\cite{ZTO PPC} while the photoresponse
was shown to be very weak at room temperature in crystalline IGZO TFTs.\cite{K.Nomura:Science03}
If NBIS is applied in $a$-IGZO TFTs, the negative $V_{\rm th}$ shift
increases up to 18 V for the gate bias of $-20$ V, and it is recovered
by annealing.\cite{Fung:JID, Takechi:JJAP09, K.Lee:APL09, ETRI-J.09}
The NBIS instability was suggested to result from the charge trapping of photo-induced
holes.\cite{Takechi:JJAP09, K.Lee:APL09, ETRI-J.09}
Under negative bias stress, free holes generated in the active region are likely to drift toward the interface
and then trapped by interfacial defects or further injected into the dielectrics.
However, the details of charge traps and their roles in the NBIS instability are not clearly understood yet.

In this letter, we report the results of first-principles calculations for the defect properties
of O-vacancy ($V_{\rm O}$) in $a$-IGZO and dielectrics and the band alignment between
the channel and dielectric materials.
We perform molecular dynamics simulations for the thermal stability of neutral and
ionized $V_{\rm O}$ defects, and find that $V_{\rm O}$ acts as a hole trap.
We suggest that V$_{\rm O}$ plays a role in the photo-excited threshold voltage instability
of oxide TFTs under negative bias stress.


We performed the density-functional calculations using the generalized gradient
approximation (GGA) for the exchange-correlation potential \cite{GGA} and the projector
augmented wave potentials,\cite{PAW} as implemented in the VASP code.\cite{VASP}
To obtain reliable valence band offsets and defect levels of $V_{\rm O}$,
we also carried out the GGA+$U$ calculations, where $U$ represents the on-site Coulomb interaction
between the localized metal $d$ electrons.
We chose the parameters, $U$ = 7, 8, and 8 eV for the In, Ga, and Zn $d$ orbitals, respectively.
For a supercell containing 84 host atoms, the wave functions were expanded in plane waves with
an energy cutoff of 400 eV.
We generated several atomic models for $a$-IGZO with the composition ratio of In:Ga:Zn = 1:1:1
through melt-and-quench molecular dynamics (MD) simulations based on the GGA,\cite{MD}
and then fully optimized the supercell volume and the ionic coordinates through the GGA+$U$ calculations.


The average volume of optimized $a$-IGZO per unit formula is about 84.8 {\AA}$^3$, 12 \% larger than
that of crystalline InGaO$_3$(ZnO)$_m$ with $m$ = 1 ($c$-IGZO).
The average coordination numbers of the In, Ga, and Zn atoms are 5.14, 4.03, and 4.17, respectively.
The GGA+$U$ band gaps of three amorphous models are 1.78, 1.74, and 1.44 eV,
as compared to the bulk value of 2.51 eV for $c$-IGZO.
The smaller band gaps result from the enlarged volume and the formation of localized states near the
valence band maximum (VBM) by disorders and under-coordinated O atoms in amorphous phase,
while the conduction band minimum (CBM) state is not much affected.
If the localized edge states are excluded, three amorphous models have similar band gaps
close to 1.78 eV.
In addition, we find the Zn 3$d$ bands to be positioned at $-$9.48, $-$9.44, and $-$9.43 eV from
the CBM, which are in good agreement with hard x-ray photoelectron spectroscopy measurements
with including a band gap correction.\cite{K.Nomura:APL08:HXPES}

In $c$-IGZO, it was reported that $V_{\rm O}$ acts as a negative-$U$ defect,
with the charge transition level $\epsilon^{+2/0}$ below the CBM.\cite{Vo in c-IGZO WJLee}
Thus, the +2 charge state of $V_{\rm O}$ is less stable than the neutral one when the electron chemical
potential is located at the CBM.
The negative-$U$ behavior of $V_{\rm O}$ is also found for $a$-IGZO.
For neutral $V_{\rm O}$ defects in $a$-IGZO, we find that their formation energies range from 2.96
to 5.82 eV, depending on the type of the neighboring metal atoms.
Among Ga$-$Ga, Ga$-$M, and M$-$M bonds around the $V_{\rm O}$ defects,
where M = In and Zn, Ga$-$Ga bonds are found to be strongest, with smaller bond lengths.
There is a tendency that $V_{\rm O}$ is easily generated at the O site surrounded by the In atoms [Fig. 1(a)],
whereas it costs a much higher energy to make an O deficiency in the environment of the Ga atoms [Fig. 1(b)].
This result agrees with the fact that adding the Ga atoms increases the stability of $a$-IGZO.\cite{T.Kamiya:JDT09}

The defect levels of $V_{\rm O}$ are also affected by the local environment at vacancy site.
We find that most $V_{\rm O}$ defects have deep donor levels, which are induced by inward relaxations
of the neighboring metal atoms.
The average defect levels are positioned at 0.81, 0.93, and 1.10 eV above the VBM for neutral
$V_{\rm O}$ defects with the Ga$-$Ga, Ga$-$M, and M$-$M bonds, respectively.
In some cases, where the neighboring metal atoms are strongly back-bonded to the second neighboring
O atoms, outward relaxations take place, resulting in the spontaneous ionization of $V_{\rm O}$, with
the ionized electrons in the conduction band [Fig. 1(c)].
Similar shallow defects were also found in previous GGA calculations for $a$-IGZO.\cite{T.Kamiya:JDT09}
If $V_{\rm O}$ captures holes, becoming the +2 charge state, the neighboring metal atoms undergo
large outward relaxations, raising the unoccupied defect level above the CBM.
In the GGA and GGA+$U$ calculations, which give the band gaps of 0.91 and 1.78 eV, the $\epsilon^{+2/0}$
levels are located at 0.86 and 0.41 eV above the CBM, respectively.
As the energy of $V_{\rm O}^{+2}$ is significantly lowered, the hole capture occurs more easily in $a$-IGZO
than in $c$-IGZO.
Here we point out that the hole trap model for $V_{\rm O}$ is consistent with the usual electron
donor model, in which $V_{\rm O}$ is a deep donor in ZnO,\cite{Janotti} although a debate still remains
because $V_{\rm O}$ is often considered as the source of electrons, based on experimental observations.

\begin{figure}
\includegraphics[clip=true,width=8.5cm]{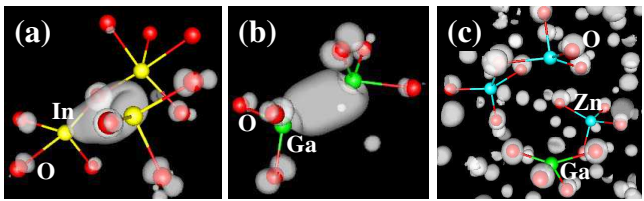}
\caption{ Isosurfaces of the charge densities for the deep levels of $V \rm _O$, with the (a) In or (b) Ga atoms
in the neighborhood, and (c) the shallow $V \rm _O$ defect surrounded by one Ga and three Zn atoms,
with electrons in the CBM. }
\label{fig1}
\end{figure}

When a device is under negative gate bias, hole carriers, which are generated in the channel by
light illumination, tend to move toward the dielectrics.
The potential barrier for hole conduction is determined by the valence band offset (VBO) between the
channel and dielectric materials.
At this point, it is difficult to make an atomic model for channel/dielectric interface structures.
Here we estimate the VBOs between $c$-IGZO, $a$-IGZO, and ZnO from their band alignments,
using the Zn 3$d$ bands as a reference level.
The GGA+$U$ calculations show that the VBM of $c$-IGZO is lower by 0.3 eV than that of ZnO,
whereas it is higher by 0.3 eV for $a$-IGZO.
We also examine a $c$-IGZO-ZnO-GaO-ZnO heterostructure, where 2 unit cells of $c$-IGZO,
8 layers of ZnO, one GaO inversion layer, and 7 layers of ZnO are stacked along the hexagonal axis.
The VBO between $c$-IGZO and ZnO agrees to within 0.2 eV with that determined from the local density
of states by removing the effect of interface dipole fields.
Considering the recent theoretical result that the VBO is nearly zero at the interface between ZnO and
amorphous HfO$_2$ ($a$-HfO$_2$),\cite{Ryu09} we finally obtain the VBO of about 0.3 eV between
$a$-IGZO and $a$-HfO$_2$, as shown in Fig. 2.
The VBOs are also compared between $a$-IGZO and other dielectric materials such as SiO$_2$
and Al$_2$O$_3$, based on the results in literatures.\cite{Ryu09, VBO of dielectrics}
As the VBOs are less than 2 eV, holes formed near the valence band edge move easily toward
the gate insulator under large negative bias stress.

\begin{figure}
\includegraphics[clip=true,width=7.5cm]{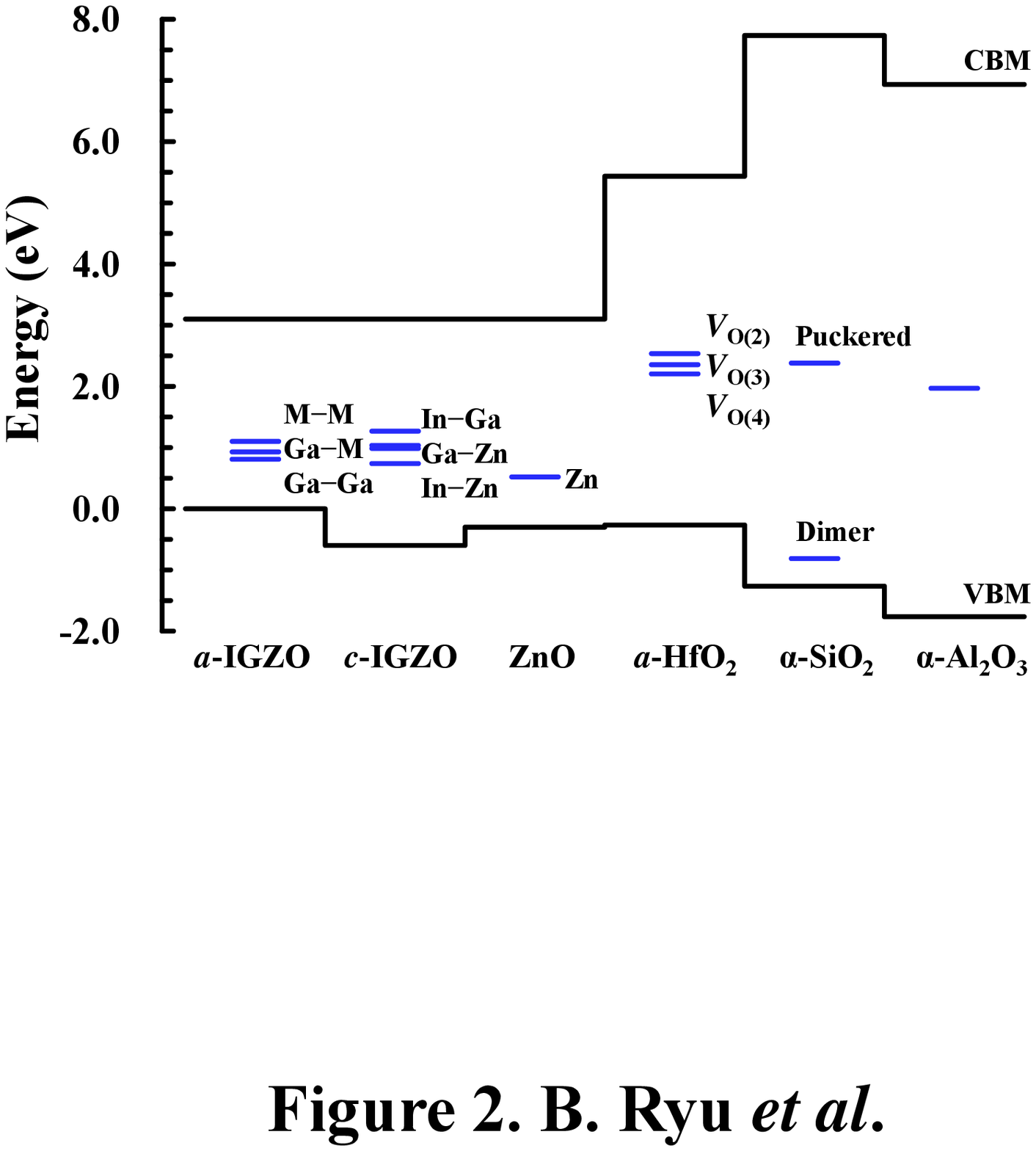}
\caption{ The defect levels of the neutral $V \rm _O$ defects in the band diagram of $a$-IGZO, $c$-IGZO, ZnO,
and dielectric oxides. The VBOs between HfO$_2$, SiO$_2$, and Al$_2$O$_3$ are from Ref. 22. }
\label{fig2}
\end{figure}

The defect levels of $V_{\rm O}$ in $a$-IGZO, ZnO, and dielectrics are compared in the band diagram
with adding the gap corrections to the CBM (Fig. 2).
Comparison of the defect levels suggest a mechanism for the NBIS degradation in $a$-IGZO TFTs.
Under illumination of UV or visible light, electron-hole pairs are mostly generated in $a$-IGZO because
the band gap of the dielectrics is larger than that of the active channel material.
Meanwhile, the $V_{\rm O}$ defects are also photo-excited, donating electrons to the conduction band.
When $V_{\rm O}$ is ionized, excited electrons give rise to persistent photoconductivity.
As the stabilization of the ionized $V_{\rm O}^{+2}$ defect is enhanced in $a$-IGZO, the photocurrent
will last longer, in good agreement with experiments.\cite{ZTO PPC}
When a large negative gate voltage is applied under light illumination [Fig. 3(a)], holes in the active region
drift toward the channel/dielectric interface due to the small VBO, whereas photo-excited electrons flow
away from the interface.
In this case, ionized $V \rm _O^{+2}$ defects can also diffuse toward the interface, resulting in the
hole trapping near the interface, as will be discussed shortly.
If holes are further injected, they will be captured by neutral $V_{\rm O}$ defects in the
dielectrics, which exhibit deep defect levels (Fig. 2).
Note that the hole capture can occur in amorphous SiO$_2$ due to the existence of various types of
$V_{\rm O}$,\cite{VO SiO2} although a dimer structure with the shallow defect level is more favorable
than a puckered configuration in crystalline $\alpha$-quartz.
The negative shift of $V_{\rm th}$ is caused by the $V_{\rm O}$ defects, which act as hole traps in
the channel and dielectrics.
After light is turned off, positively charged $V_{\rm O}^{+2}$ defects accumulated near the interface
cause a band bending between $a$-IGZO and dielectrics, forming a potential well, as shown in Fig. 3(b).
Due to electron carriers confined in the well, devices can exhibit finite currents in the absence of
gate voltage, so that a negative voltage is needed to make the current zero.

\begin{figure}
\includegraphics[clip=true,width=8.5cm]{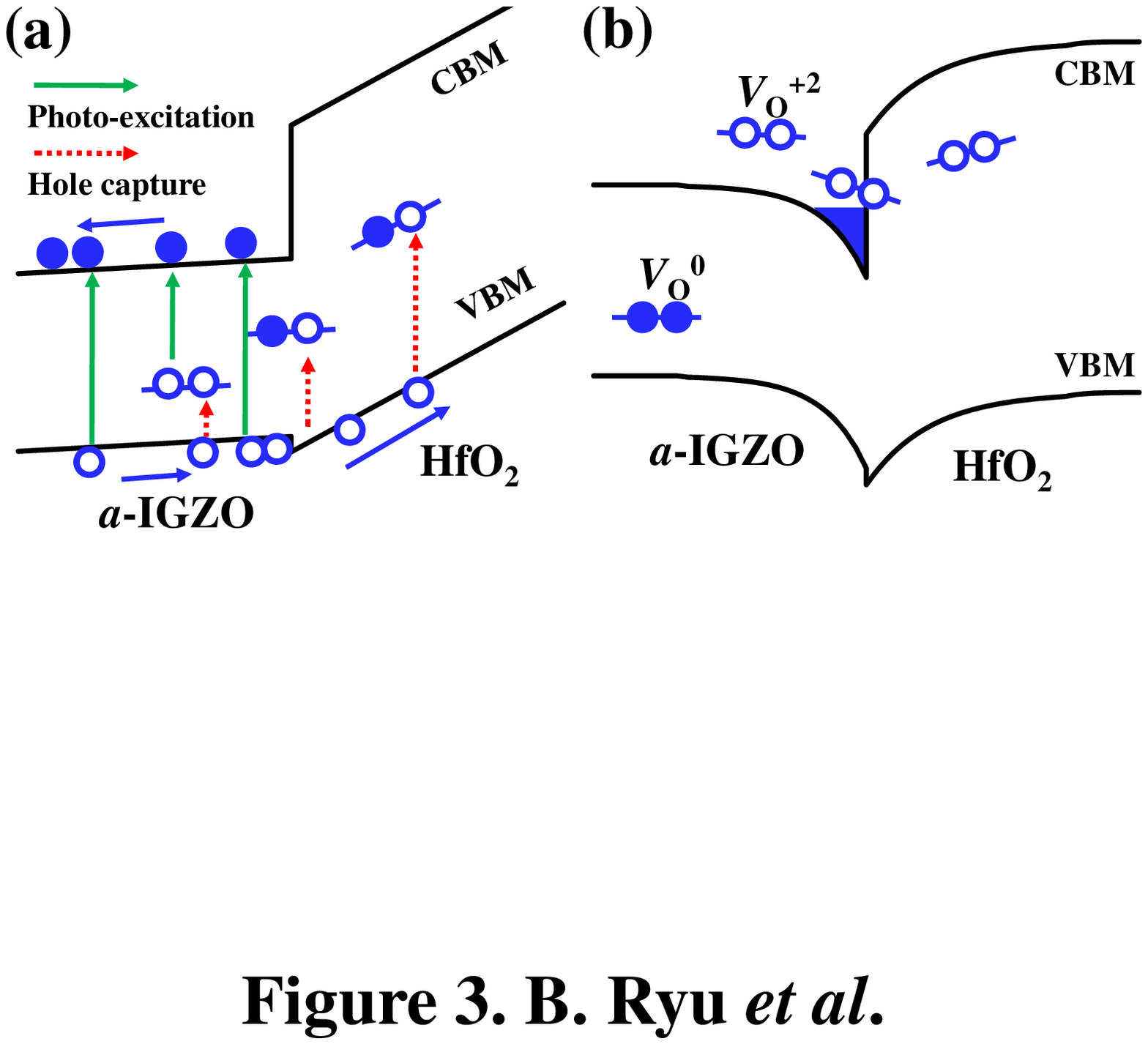}
\caption{ Schematic illustrations of the photo-excitation of electron-hole pairs and the $V \rm _O$ defects,
the drift motion of hole carriers, the hole trapping in the channel/dielectric interface (a) under the NBIS and
(b) after swithing off the NBIS. In (b), a potential well is developed due to the band bending at the interface. }
\label{fig3}
\end{figure}

The hole traps in $a$-IGZO can be removed by thermal annealing.
We perform {\it ab initio} MD simulations for 8 ps at temperatures of $200 - 400$ ${\rm ^o}$C to examine
the annealing effect on the stability of $V_{\rm O}^{+2}$.
In $c$-IGZO, when $V_{\rm O}^{+2}$ captures two electrons, this defect easily returns to the original
neutral configuration.
However, in $a$-IGZO, we note that electron capture does not take place at zero temperature
for some of the $V_{\rm O}^{+2}$ defects, with keeping electrons in the conduction band.
MD simulations show three distinct annealing effects in amorphous phase. (i) Structure recovery:
$V \rm _O^{+2}$ captures electrons and returns to the original defect at the same vacancy site
even at room temperature, similar to $c$-IGZO.
(ii) Reconstruction and diffusion: In the environment of the Ga atoms, a neighboring O atom fills in the
vacancy site via reconstruction, resulting in the formation of a different $V_{\rm O}$ defect and thereby
the diffusion of $V_{\rm O}$, especially toward the O site surrounded by the In atoms.
We note that $V_{\rm O}$ can diffuse at room temperature, thus, hole traps can segregate to
the interface under negative gate bias stress.
When the localized holes are captured by $V_{\rm O}$ in the dielectrics, they can be removed by
tunneling electrons through the interface if positive bias temperature stress is applied.
(iii) Shallow level formation: A certain $V_{\rm O}^{+2}$ defect maintains the shallow level by creating a void
around the vacancy site.
In this case, the electron capture hardly occurs even at the annealing temperature of $500 - 600$ ${\rm ^o}$C,
higher than the crystallization temperature.\cite{T.Kamiya:JDT09}
It is clear that the NBIS instability can be recovered by the annealing effects in (i) and (ii).
However, regarding the stable ionized defect in (iii), we point out that the GGA+$U$ may overestimate the
annealing temperature for the recovery of the neutral deep state.
In the GGA+$U$, the energy barrier between the deep neutral and ionized shallow states seems to
be overestimated because the $\epsilon^{+2/0}$ level appears above the CBM, as mentioned earlier.
If the GGA and GGA+$U$ results for the band gap are extrapolated to the measured band gap
of 3.1 eV,\cite{Janotti} the extrapolated value of $\epsilon^{+2/0}$ will be
within the band gap, lying at 2.85 eV above the VBM.
With including the band gap correction, it is expected that the annealing temperature will be reduced
in MD simulations.

In conclusion, we have shown that the $V_{\rm O}$ defects in the channel and dielectric materials
play a role in the $V_{\rm th}$ instability of oxide TFTs under negative bias illumination stress.
We have proposed a model, where photo-excited holes in the active channel are either captured
by the $V_{\rm O}$ defects or drift toward the interface due to the small potential barrier
for hole conduction.
The injected hole carriers through the interface can be also captured by the $V \rm _O$ defects
in the dielectrics under large negative bias stress, causing the negative shift of $V_{\rm th}$.
Due to the amorphous structure of IGZO, some of the $V \rm _O^{+2}$ defects, which trap holes,
hardly recover their original defect states by electron capture, maintaining the shallow defect levels.
The MD simulations have shown that the hole traps can be removed by thermal annealing, recovering
the device stability.

This work was supported by National Research Foundation of Korea (Grant No. NRF-2009-0093845)
and Information and Technology R\&D program (2006-S079-02) of the Ministry of Knowledge Economy.

\end{document}